\documentclass[aps,twocolumn,pra,superscriptaddress,amsmath,showpacs,tightenlines]{revtex4-1}
\usepackage{amssymb}
\usepackage{amsmath}
\usepackage{graphicx}
\usepackage{subfigure}
\usepackage{natbib}
\usepackage{epsfig}
\usepackage{amsfonts}
\usepackage{mathrsfs}
\usepackage{xcolor}
\usepackage[toc,page,title,titletoc,header]{appendix}
\begin{document}

\title{{Single-photon scattering and bound states in an atom-waveguide system with two or multiple coupling points}}

\author{Wei Zhao}
\affiliation{Center for Quantum Sciences and School of Physics, Northeast Normal University, Changchun 130024, China}
\author{Zhihai Wang}
\email{wangzh761@nenu.edu.cn}
\affiliation{Center for Quantum Sciences and School of Physics, Northeast Normal University, Changchun 130024, China}

\begin{abstract}
In this paper, we investigate the single-photon scattering and bound states in a one-dimensional coupled-resonator waveguide which couples to a single {artificial giant atom with two or more coupling points}. When the atom couples to the waveguide via two resonators, the single-photon  reflection rate is characterized by either Breit-Wigner or Fano line shapes. When the atom couples to the waveguide via multiple resonators, we numerically show how the destructive interference effect leads to a complete single-photon reflection. We also find a phase transition phenomena for the multi-resonator coupling case,  which reveals that the upper bound state only exists when the atom-waveguide coupling strength is above a critical value.
\end{abstract}

%\date{\today}

\maketitle
\section{introduction}
The light-matter interaction is a central topic in the field of quantum optics. Recent years, much attention has been focused on the light-matter interaction in waveguiding structures, leading to a scenario named as waveguide QED. As reviewed in Refs.~\cite{DR2017,XG2017} and the references therein, there are lots of theoretical and experimental works on waveguide QED system, for example, the single-photon device~\cite{JT2005,DE,LZ2008}, the phase transition~\cite{MF2017,LQ2019}, dressed or bound states~\cite{HZ,GC2016,TS2016,ES2017,PT2017,GC2019},  the exotic topological and chiral phenomena~\cite{MR2014,VY,IM,CG2016,MB2019,SM2019}, just name a few.

In the sense of quantum network~\cite{HJ2008}, the waveguide is usually regarded as quantum channel for photons, and the atom (or artificial atom) plays as quantum node. One of the subject in waveguide QED is how to control the propagation of the photons in quantum channel by adjusting the quantum node(s). In the traditional scheme, the size of the atom is at least one order smaller than the wavelength of the propagating photons in the waveguide, therefore it is reasonable to approximate the atom as a point-like dipole. Recently, a superconducting transmon qubit was successfully coupled
to propagating surface acoustic waves~\cite{SDbook,DM,MV2014,RM2017}. Due to the slow propagation speed of sound in solids, the wavelength of the phonons for a given frequency can be smaller than the size of the atom, and the point-like dipole approximation for the atom does not work. In this situation, we must deal with a {``giant atom''}  setup~\cite{TP2003,AF2014,LG2017,AF2018,PT,GA2019,AG2019,LG2019,SG2019,BK2019}, in which the size of the atom provides us another controller besides the resonant frequency and the dipole moment, for the states of the photons in the waveguide.

In this paper, we investigate the single-photon scattering and bound states in a one-dimensional coupled-resonator waveguide with {giant atom}, which can be coupled to the waveguide via two or multiple resonators. For the two-resonator coupling situation, we analytically obtain the single-photon scattering behavior and find that the Breit-Wigner or Fano~\cite{UF1961} line shapes for the reflection rate take turns as the size of the {giant atom} changes. For multiple-resonator coupling situation, we numerically demonstrate the destructive interference, which finally leads to the complete single-photon reflection. Besides, we find a phase transition phenomena when the {giant atom} couples to the waveguide via multiple resonators. That is, when the atom-waveguide coupling strength surpasses a certain value, there will be two bound states asymmetrically located above and below the propagating band, otherwise, there will be only one, which locates below the propagating band.

The rest of the paper is organized as follows. In Sec.~\ref{model}, we present our model and the Hamiltonian. In Sec.~\ref{tscattering} and \ref{mscattering}, we study the single-photon scattering when the {giant atom} couples to the waveguide via two and multiple resonators, respectively. In Sec.~\ref{bound}, we discuss the properties of the bound states and end up with a brief conclusion in Sec.~\ref{con}. Some detailed derivation in the momentum space is given in the Appendix.

\section{Model and Hamiltonian}
\label{model}
{As schematically shown in Fig.~1, the system we consider is composed by a
one-dimensional coupled-resonator waveguide with infinite length and a two-level system.} In the conventional photonic waveguide scenario, the natural atom (for example, the Rydberg atom) only coupled to a single resonator due to its small size. However, we now study the effect of a giant atom scheme, where the two-level system can be coupled to the waveguide via two or more resonators simultaneously.
\begin{figure}
\centering
\includegraphics[width=0.5\textwidth]{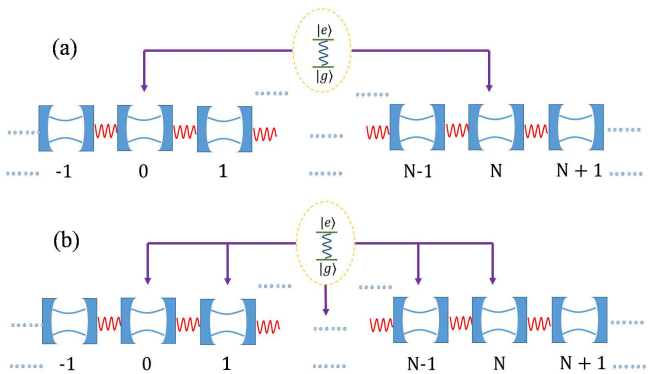}
\caption{{Schematic configuration for a one-dimensional coupled-resonator waveguide coupled to a giant atom. (a) The giant atom couples to two resonators. (b) The giant atom couples to multiple resonators.}}
\label{scheme}
\end{figure}

The coupled-resonator waveguide is modelled by the Hamiltonian
\begin{equation}
{H_{c}=\omega_c\sum_{j}a_{j}^{\dagger}a_{j}-\xi\sum_{j=-\infty}^{\infty}\left(a_{j+1}^{\dagger}a_{j}+a_{j}^{\dagger}a_{j+1}\right)},
\end{equation}
where $\omega_c$ is the frequency of the resonators, and $a_j$ is the bosonic annihilation operator on site $j$. $\xi$ is the hopping strength between the nearest resonators.

We now formulate the interaction between the {giant atom} and the waveguide. In this paper, we will discuss the single-photon scattering and bound states in two kinds of setups as follows. {In case (I), as shown in Fig.~\ref{scheme}(a), the atom only couples to the $0$th and the $N$th resonators with the same coupling strength $J$}, the Hamiltonian for the whole system is written as
\begin{equation}
H_1=H_c+\Omega|e\rangle \langle e|+J[(a_{0}^{\dagger}+a_{N}^{\dagger})\sigma^{-}+{\rm H.c.}]
\label{H1}
\end{equation}
where $\sigma^{\pm}$ are the usual Pauli operators of the {giant atom}, and $\Omega$ is the transition frequency between the ground state $|g\rangle $ and the excited state $|e\rangle$. {In case (II), as shown in Fig.~\ref{scheme}(b),} the atom uniformly couples to $N+1$ resonators with the resonator number $j=0\rightarrow N$ simultaneously. The Hamiltonian is expressed as
\begin{equation}
{H_2=H_c+\Omega|e\rangle \langle e|+\frac{2J}{N+1}\sum_{j=0}^{N}\left(a_{j}^{\dagger}\sigma^{-}+{\rm H.c.}\right)}.
\label{H2}
\end{equation}
We note that, the total coupling strength between the giant atom and the waveguide are both $2J$ for the two cases. Here, we have performed the rotating wave approximation, which is valid in the parameter regime $J\ll\Omega$ and $\Omega\sim\omega_c$.

{In Ref.~\cite{AF2018}, the coupling between a giant atom and linear waveguide has been systematically studied. Here, we propose a  discrete version, where the waveguide provide a structured environment for the atom.  The giant atom here means that the atom can be coupled to two or more desirable resonators. Compared to the linear waveguide in the literatures, the coupled-resonator waveguide supplies us a propagating channel in which the photonic velocity can be tuned by adjusting the inter-resonator coupling strength. Meanwhile, the coupled-resonator waveguide form an energy band, which is centered at $\omega_c$ and the total width is $4\xi$ and the dressed atom will contribute to the bound state outside the band, as shown in the following sections.}

\section{Single-photon scattering with two coupling resonators}
\label{tscattering}
In this section, we will study the single-photon scattering in case (I), that is, the giant atom only couples to the $0$th and $N$th resonators in the waveguide. We now consider that a single photon with wave vector $k$ is incident from the left side of the waveguide. Since the excitation number in the system conserves, the eigenstate in the single-excitation subspace can be written as
\begin{equation}
|E_k\rangle=\left(\sum_j c_j a_j^{\dagger}+u_e\sigma^{+}\right)|G\rangle,
\label{EK}
\end{equation}
where $|G\rangle$ represents that all of the resonators in the waveguide are in the vacuum states, while the giant atom is in the ground state $|g\rangle$. $c_j$ is the probability amplitude for finding a photonic excitation in resonator $j$, and $u_e$ is the excitation amplitude of the {giant atom}. In the regime $j<0$ and $j>N$, the amplitude $c_j$ can be written in the form of
\begin{equation}
c_{j}=\begin{cases}
e^{ikj}+re^{-ikj} & j<0\\
te^{ikj} & j>N
\end{cases},
\label{tail}
\end{equation}
where $r$ and $t$ are respectively the single-photon reflection and transmission amplitudes. Hereafter, the wave vector $k$ is considered to be dimensionless by setting the distance between two arbitrary neighboring resonators as unit. In the regime covered by the {atom}, the photon propagates back and forth, and the amplitude $c_j$ for $0\leq j\leq N$ can be expressed as
\begin{equation}
c_j=A e^{ikj}+Be^{-ikj}.
\label{inside}
\end{equation}
The Sch\"{o}dinger equation $H_1|E_k\rangle=E_k|E_k\rangle$ yields the dispersion relation $E_k=\omega_c-2\xi\cos k$. Together with the continuous condition at $j=0$ and $j=N$, which are $1+r=A+B$ and $A e^{ikN}+Be^{-ikN}=te^{ikN}$, respectively, the reflection rate $R=|r|^2$ can be obtained as
\begin{equation}
R=\frac{4J^{4}\cos^{4}\left(\frac{kN}{2}\right)}{4J^{4}\cos^{4}\left(\frac{kN}{2}\right)
+\left[\xi\Delta_k\sin k-J^{2}\sin\left(kN\right)\right]^{2}},
\label{reflection}
\end{equation}
where $\Delta_k=E_k-\Omega$ is the detuning between the {atom} and the propagating photons in the waveguide.

\begin{figure}
\centering
\includegraphics[width=0.85\columnwidth]{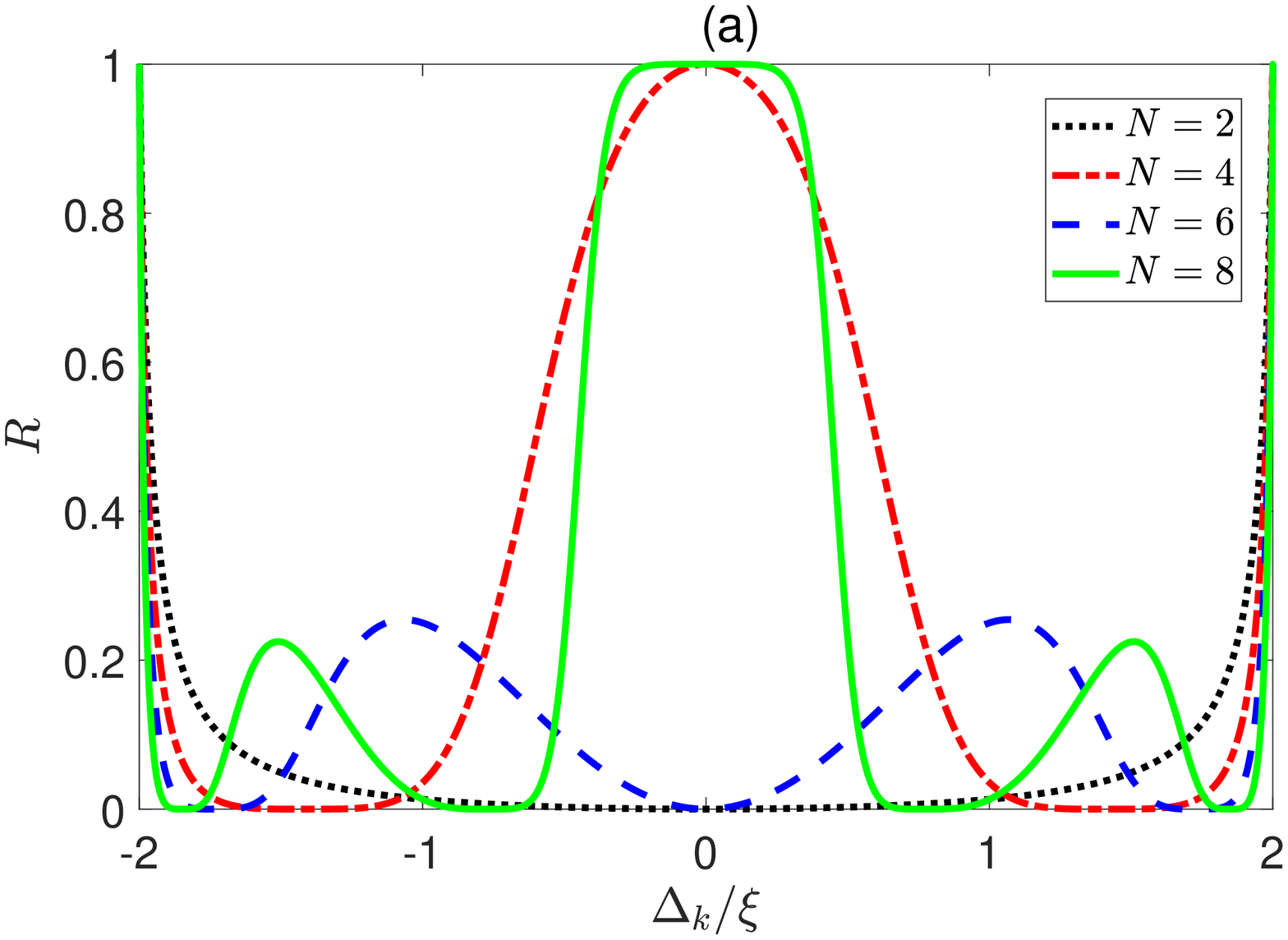}
\includegraphics[width=0.85\columnwidth]{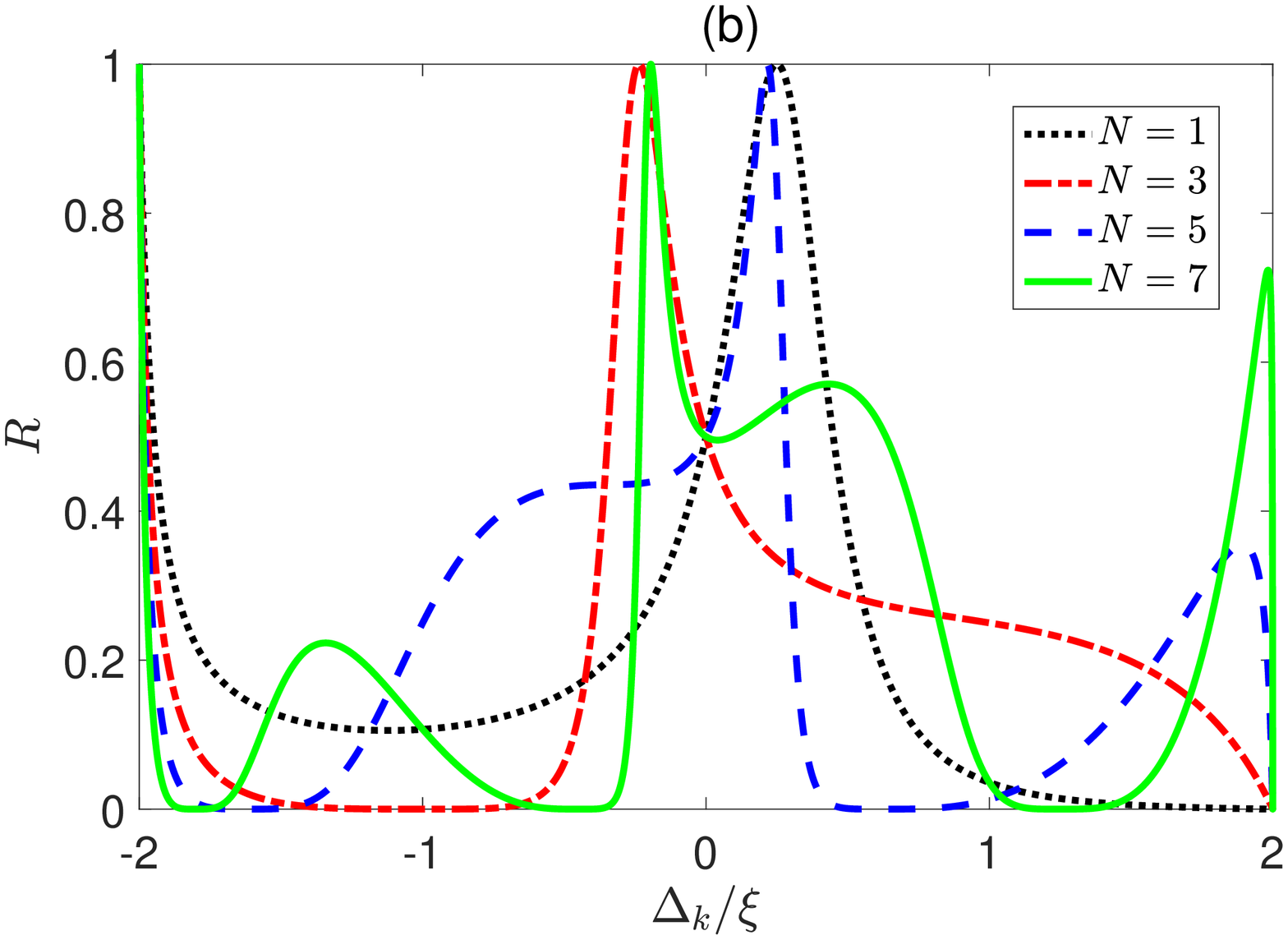}
\includegraphics[width=0.85\columnwidth]{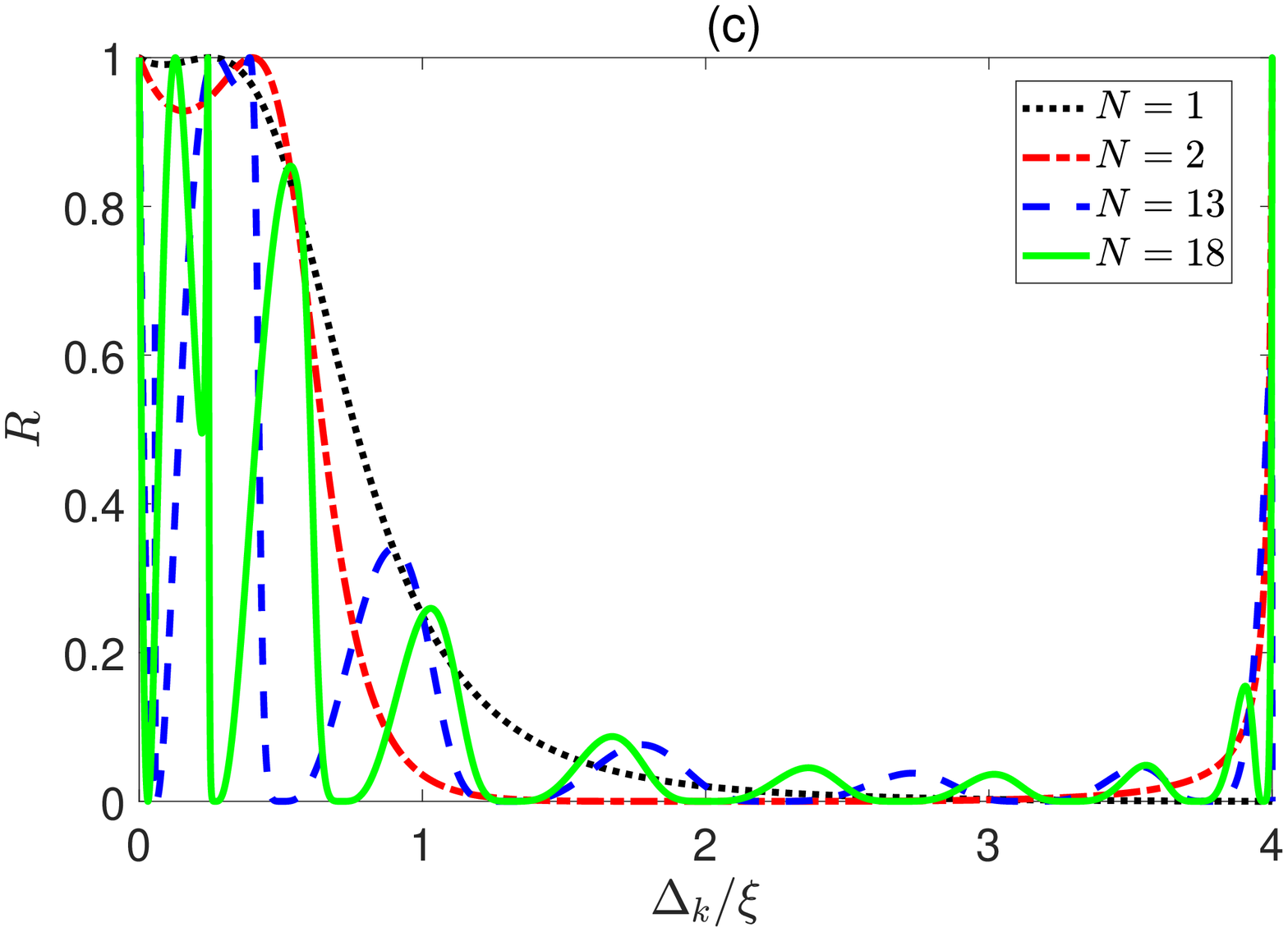}
\caption{The reflection rate $R$ as a function of the detuning $\Delta_k$ for different $N$ with two coupling resonators. The parameters are set as $J=0.5\xi,\omega_c=20\xi$. $\Omega=20\xi$ for (a) and (b), and $\Omega=18\xi$ for (c).}
\label{Rreflect}
\end{figure}

In the small atom scenario ($N=0$), we recover the results given in Ref.~\cite{LZ2008}, and it is obvious that the incident photon will be completely reflected ($R=1$) when it is resonant with the atom, that is, $\Delta_k=0$. However, for the {giant atom} situation under our consideration, the photon can be reflected by the two connecting points, the scattering process will be dramatically affected by the size of the {giant atom} . In Fig.~\ref{Rreflect}, we plot the reflection rate $R$ as a function of the detuning for different $N$, which characterizes the size of the {giant atom} .

Let us first discuss the situation of $\omega_c=\Omega$. In this case, the giant atom is resonant to the propagating photons with wave vector $k=\pi/2$. In Fig.~\ref{Rreflect}(a) and (b), we demonstrate the results for even and odd $N$, respectively. For even $N$ in Fig.~\ref{Rreflect}(a), at the resonant frequency, $\Delta_k=0$, the incident photon will be completely transmitted ($R=0$) for $N=4m+2$ and completely reflected ($R=1$) for $N=4m\,(m=0,1,2\cdots)$. Furthermore, the curve always represents a Breit-Wigner-like line shape around the resonance $\Delta_k=0$, which is similar to the small atom case {as shown in Fig.~\ref{Rreflect}(a) by the green solid line. However, the width of the reflection window is much larger for the giant atom ($N\neq0$)}.  Meanwhile, for odd $N$, we observe a frequency shift in Fig.~\ref{Rreflect}(b), with the complete reflection occurring at a positive (negative) detuning for $N=4m+1(4m+3)$, and the curve behaves as a Fano-like shape around the peak, which is not for small atom situation. Moreover, we can also find that $R=0.5$ for $\Delta_k=0$, which is independent of the size of {giant atom}.

Next, we consider the situation of $\omega_c\neq\Omega$. As shown in Fig.~\ref{Rreflect}(c), the reflection rate shows a complicated dependence on the detuning $\Delta_k$. From Eq.~(\ref{reflection}), the detuning for complete reflection is determined by the
transcendental equation
\begin{equation}
\Delta_k=\frac{J^{2}\sin\left(kN\right)}{\xi\sin k},
\end{equation}
around which, the reflection yields a Fano shape. We also observe that, there will be one or more complete reflection frequencies (excluding the edge of the photonic propagating bands), depending on the values of $N$. It implies that, we can design the on-demand single-photon transistor by adjusting the size of {giant atom in our waveguide setup. On the contrary, for the case of small atom, as shown in Fig.~\ref{Rreflect}(c) for $N=0$, the complete reflection only occurs at $\Delta_k=0$}.

\section{Single-photon scattering with multiple coupling resonators}
\label{mscattering}

Now, we move to case (II), where the  {giant atom} couples to all of the resonators it covers, and the Hamiltonian of the system is described by Eq.~(\ref{H2}). In this case, we will give a semi-analytical result for the single-photon scattering behavior.

{We note that the Hamiltonian in Eq.~(\ref{H2}) can be rewritten as
$H_2=H_L+H_R+H_s+H_{\rm Int}$, where
\begin{eqnarray}
H_{L}&=&\sum_{j=-\infty}^{-1}\omega_c a_{j}^{\dagger}a_{j}-\xi\sum_{j=-\infty}^{-2}(a_{j}^{\dagger}a_{j+1}
+a_{j+1}^{\dagger}a_{j}), \\
H_{R}&=&\sum_{j=N+1}^{\infty}\omega_c a_{j}^{\dagger}a_{j}-\xi\sum_{j=N+1}^{\infty}
(a_{j}^{\dagger}a_{j+1}+a_{j+1}^{\dagger}a_{j}), \\
H_s&=&\Omega|e\rangle \langle e|+\omega_c\sum_{j=0}^N a_j^\dagger a_j+\frac{2J}{N+1}\sum_{j=0}^{N}\left(a_{j}^{\dagger}\sigma^{-}+{\rm H.c.}\right)\nonumber \\&&-\xi\sum_{j=0}^{N-1}(a_{j}^{\dagger}a_{j+1}+a_{j+1}^{\dagger}a_{j})\nonumber \\
&=&\sum_{m}v_m|\phi_m\rangle\langle \phi_m|,\\
H_{\rm Int}	&=&	-\xi\sum_{m}(a_{-1}^{\dagger}x_{m}|0\rangle\langle\phi_{m}|
+a_{N+1}^{\dagger}y_{m}|0\rangle\langle\phi_{m}|+{\rm H.c.}),\nonumber \\
\end{eqnarray}}
Here, $v_m\,(m=1,2,\cdots N+2)$ is the $m$th {eigenvalue} of $H_s$ in the single-excitation subspace and $|\phi_m\rangle$ is the corresponding {eigenstate. We have defined $x_{m}:=\langle0|a_{0}|\phi_{m}\rangle,\, y_{m}:=\langle0|a_{N}|\phi_{m}\rangle$, where $|0\rangle$ represents that all of the resonators are in the vacuum state.}
As before, we assume that a single photon with wave vector $k$ is incident from the left side of the waveguide, the wave function in the single-excitation subspace follows
\begin{equation}
|\psi_{k}\rangle=\sum_{j=-\infty}^{-1}U_{j}a_{j}^{\dagger}
|0\rangle+\sum_{j=N+1}^{\infty}V_{j}a_{j}^{\dagger}|0\rangle
+\sum_{m}A_{m}|\phi_{m}\rangle,
\end{equation}
where $U_{j}=e^{ikj}+re^{-ikj},V_{j}=te^{ikj}$, with $r$ and $t$ being the reflection and transmission amplitudes.  The Sch\"{o}dinger equation $H_2|\psi_{k}\rangle=E_{k}|\psi_{k}\rangle$ yields $E_{k}=\omega_c-2\xi\cos k$ and
\begin{eqnarray}
(E_{k}-\omega_c)U_{-1}&=&-\xi\sum_{m}x_{m}A_{m}-\xi U_{-2},\label{RR0}\\
(E_{k}-\omega_c)V_{N+1}&=&-\xi\sum_{m}y_{m}A_{m}-\xi V_{N+2},\label{RR1}\\
(E_{k}-v_{m})A_{m}&=&-\xi\left(U_{-1}x_{m}^{*}+V_{N+1}y_{m}^{*}\right).\label{RR2}
\end{eqnarray}
Even without the detailed calculation about the scattering behavior, we can obtain the condition for a complete reflection ($t=0$) from Eq.~(\ref{RR1}), which yields
\begin{equation}
\sum_{m}y_{m}A_{m}=0.
\label{inte}
\end{equation}
It implies that the incident photon will interact with all of the modes of $H_s$, and the destructive interference will lead to a complete reflection. The similar mechanism was also found in a super-cavity scheme under the two-mode approximation~\cite{WZ2014}.
{We would like to point that, the destructive interference implied by Eq.~(\ref{inte}) is not a novel feature in giant atom coupling, it is also true for the small atom scenario. For the small atom, $|\phi_m\rangle$ ($m=1,2$) is actually the dressed states for single atom-resonator system, which can be described by the so called ``Jaynes-Cummings'' model.}

\begin{figure}
\begin{centering}
\includegraphics[width=1\columnwidth]{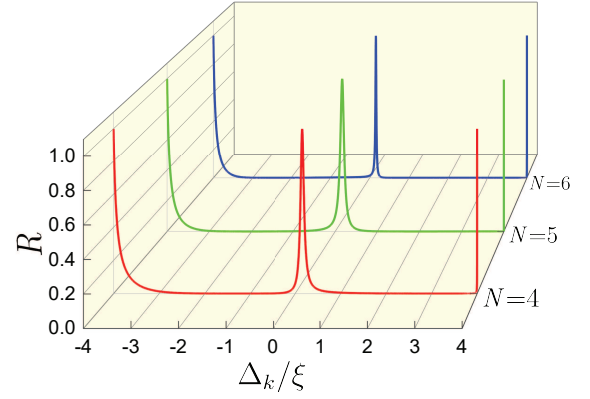}
\par\end{centering}
\caption{The reflection rate $R$ as a function of the detuning $\Delta_k$ for different $N$ with multiple coupling resonators. The parameters are set as $J=0.5\xi,\omega_c=\Omega=20\xi$.}
\label{Rreflectm}
\end{figure}

Furthermore, the single-photon scattering behavior can be predicted by solving Eqs.~(\ref{RR0}),(\ref{RR1}) and (\ref{RR2}), and the reflection rate $R=|r|^2$ is obtained as
\begin{equation}
R=\frac{(Q_{1}Q_{2}+\xi^{2}\sin^{2}k-|M_{3}|^{2})^{2}+\xi^{2}(Q_{1}-Q_{2})^{2}
\sin^{2}k}{(Q_{1}Q_{2}-\xi^{2}\sin^{2}k-|M_{3}|^{2})^{2}
+\xi^{2}(Q_{1}+Q_{2})^{2}\sin^{2}k},
\end{equation}
where $Q_{n}=E_{k}-\omega_c-M_{n}+\xi\cos k$ for $n=1,2$ and
\begin{eqnarray}
M_{1}&=&\xi^{2}\sum_{m}\frac{|x_{m}|^{2}}{E_{k}-v_{m}},\\ M_{2}&=&\xi^{2}\sum_{m}\frac{|y_{m}|^{2}}{E_{k}-v_{m}},\\
M_{3}&=&\xi^{2}\sum_{m}\frac{x_{m}y_{m}^{*}}{E_{k}-v_{m}},
\end{eqnarray}
and the condition for complete reflection is given by $|M_3|=0$.
{In Fig.~\ref{Rreflectm}, we plot $R$ as a function of $\Delta_k$ for different
relative small $N$. Different from the case of two connecting points, for multiple points $N>2$,
the complete reflection only occurs near the resonant point $\Delta_k=0$. }

\section{Bound states}
\label{bound}
\begin{figure}
\begin{centering}
\includegraphics[width=\columnwidth]{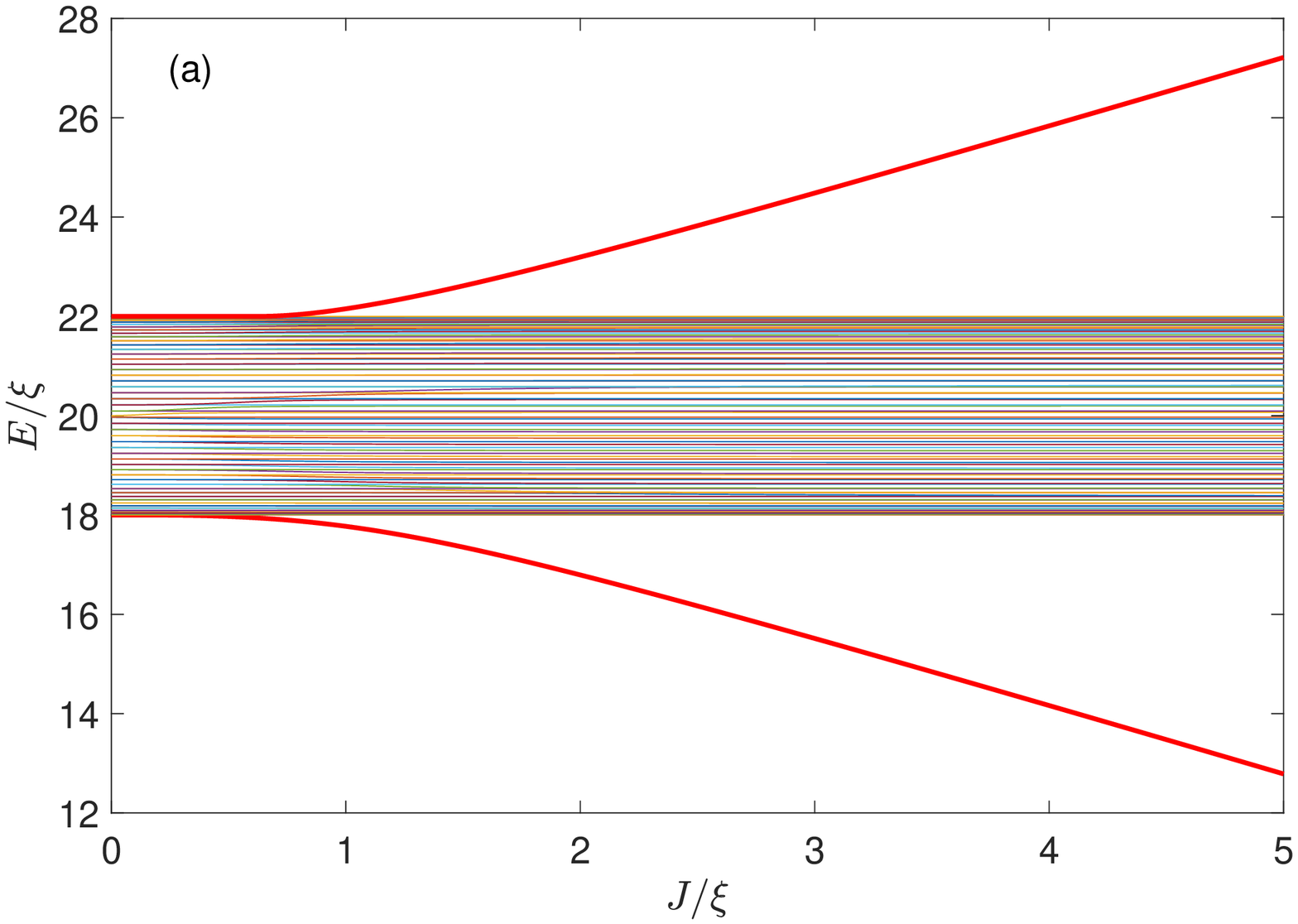}
\includegraphics[width=\columnwidth]{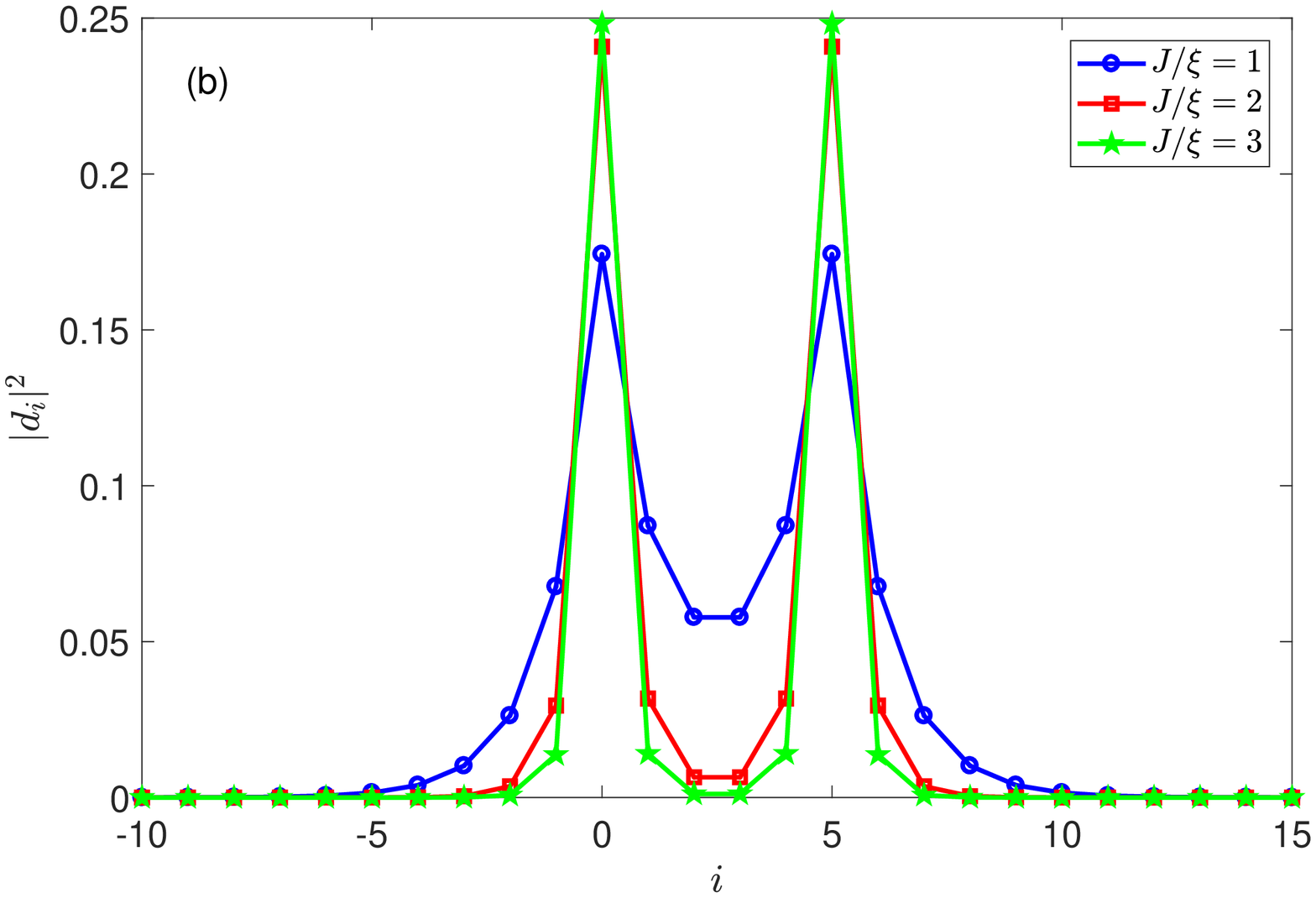}
\par\end{centering}
\caption{{The energy spectrum (a) and the photonic probability distribution of the bound state (b) in Case (I). The parameters are set as $\omega_c=\Omega=20\xi, N=5$.}}
\label{twobound}
\end{figure}
In the above sections, we have investigated the scattering states in the single-excitation subspace, with the eigen-energy inside the band $\omega_k\in[\omega_c-2\xi,\omega_c+2\xi]$. Besides, the interaction between the
{giant atom} and the waveguide breaks the translational symmetry of the waveguide, leading to another kind of bound state, where the photonic excitation is bounded near the
regime of the {atom}, and the corresponding energies lay outside the propagating band.

\begin{figure}
\begin{centering}
\includegraphics[width=\columnwidth]{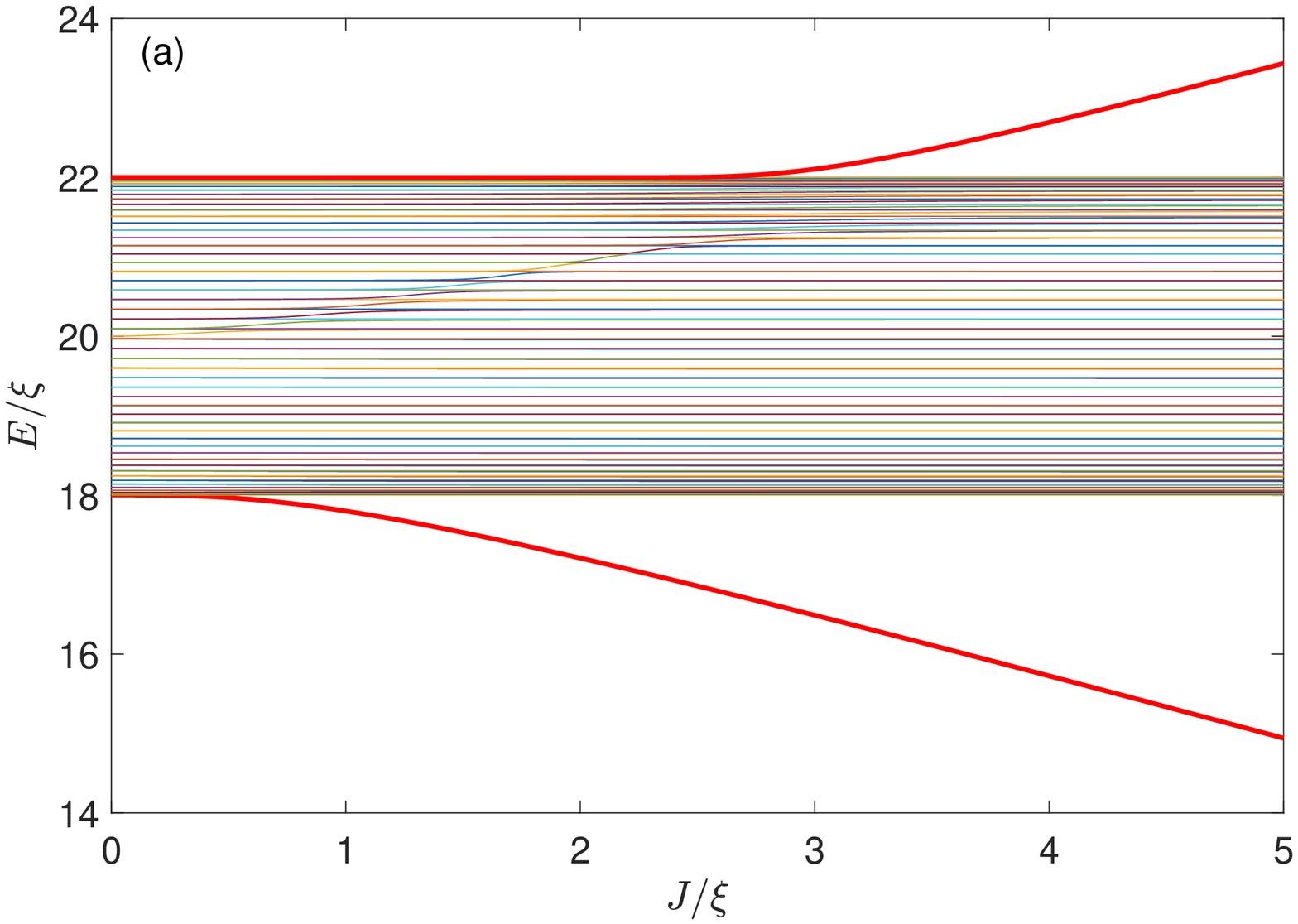}
\includegraphics[width=\columnwidth]{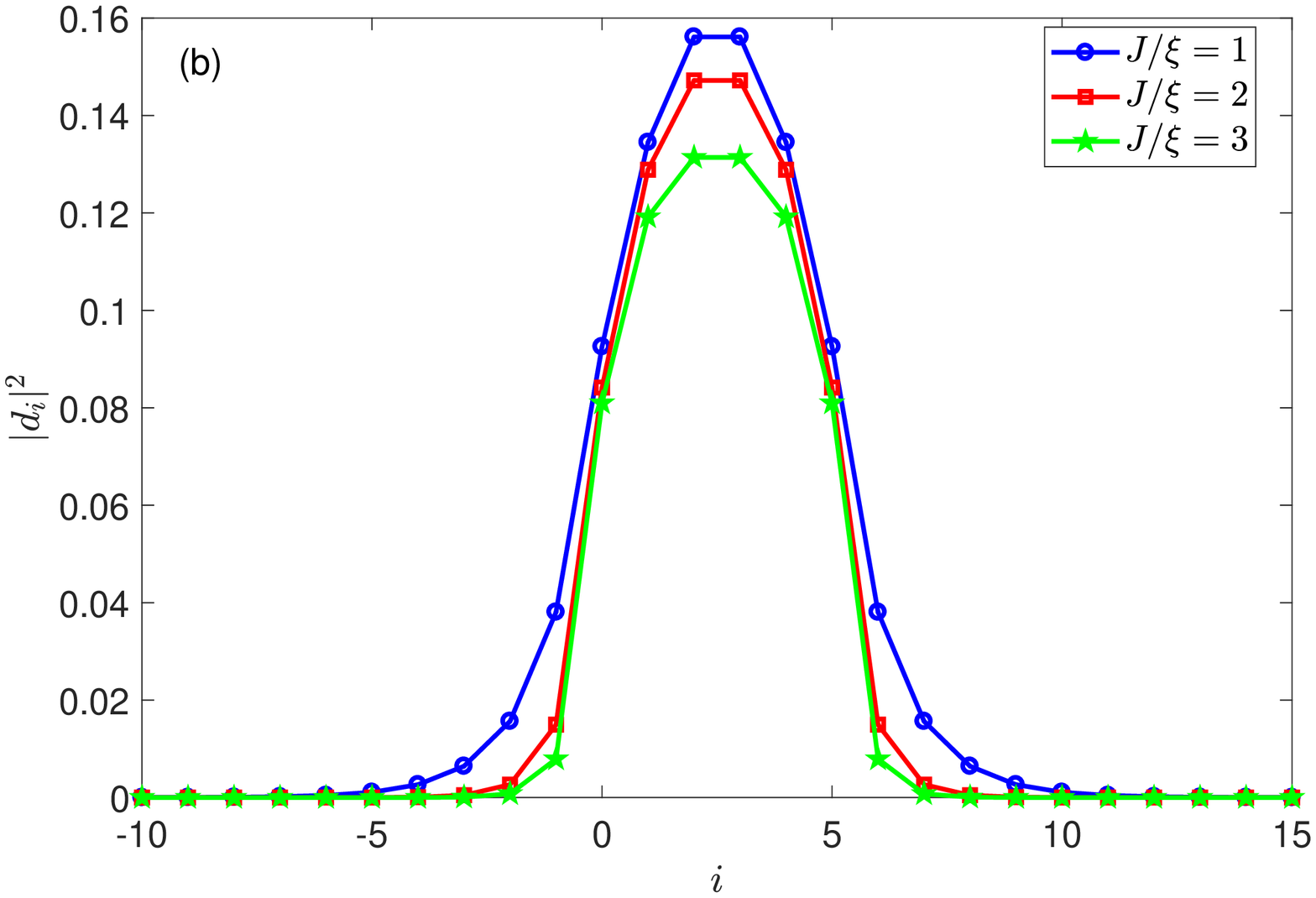}
\includegraphics[width=\columnwidth]{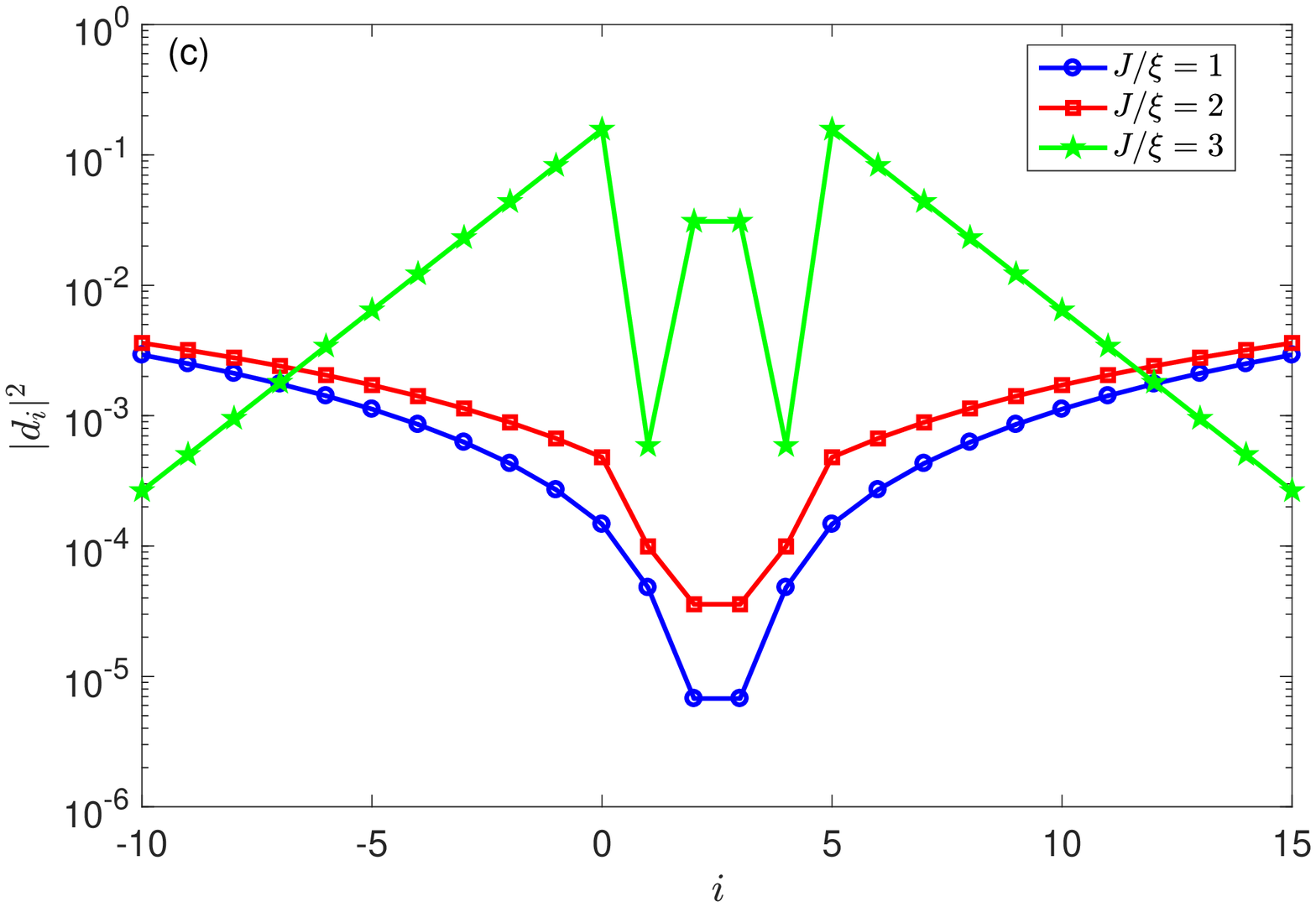}
\par\end{centering}
\caption{{The energy spectrum (a) and the photonic  probability distribution of the lower state (b) and upper state (c) in energy  for Case (II). The parameters are same with Fig.~\ref{twobound}. }}
\label{multibound}
\end{figure}

\begin{figure}
\begin{centering}
\includegraphics[width=\columnwidth]{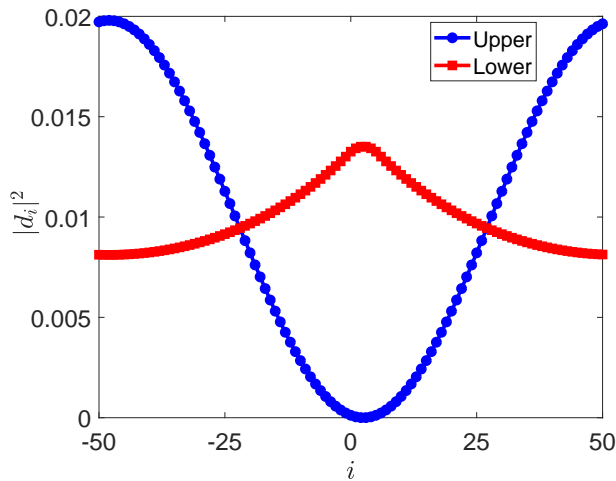}
\par\end{centering}
\caption{{The photonic  probability distribution of the lower and upper states in energy  for Case (II). The parameters are set as $J/\xi=0.1$ and the other parameters are set to be same with Fig.~\ref{twobound}. }}
\label{small}
\end{figure}

In principle, the energies for the bound state(s) can be obtained by transforming to the momentum space and solving the transcendental equation. Similar to the approach given in Ref.~\cite{GC2016}, after the detailed calculations as shown in the Appendix, the transcendental equations for the energy $E$ are
\begin{equation}
{E-\Omega=\frac{J^{2}}{\pi}\int_{-\pi}^{\pi}dk\frac{1+\cos(kN)}{E-\omega_c+2\xi\cos(k)}}
\label{eqtwobound}
\end{equation}
 and
\begin{equation}
{E-\Omega=\frac{2J^{2}}{(N+1)^{2}\pi}\int_{-\pi}^{\pi}dk\frac{\sin^{2}
[\frac{k(N+1)}{2}]}{\sin^{2}(\frac{k}{2})[E-\omega_c+2\xi\cos(k)]}}
\label{eqmbound}
\end{equation}
for Case (I) and Case (II), respectively. However, the integrals in the above two equations are not easy to be performed. Therefore, we will resort to numerical diagonalization of the Hamiltonian in the real space, and plot the single-excitation energy spectrum and the photonic distribution of the upper and lower states [the wave-function is expressed as $|E\rangle=(\sum_i d_ia_i^\dagger+d_e\sigma^+)|G\rangle$] in {Fig.~\ref{twobound}, Fig.~\ref{multibound} and Fig.~\ref{small} under the periodical boundary condition.}

In Fig.~\ref{twobound}(a), we plot the energy spectrum in the single-excitation subspace for Case (I), where the {giant atom} couples to the waveguide via two resonators. The energy band in the middle with band width $4\xi$ are the
scattering states, which are discussed in Sec.~\ref{tscattering}. The other two red curves, which are nearly symmetrically located above and {below} the propagating band are the bound states. The {photonic probability distributions} for these two states are nearly same, and are plotted in Fig.~\ref{twobound} (b). It shows that the photon is bounded around the two resonators which couple to the {giant atom}, and exponentially decays in the regime far away from the {atom}.

{As for Case (II) with $N\geq1$, where the giant atom} couples to all of the resonators it covers, the upper and lower energy levels are not symmetric about the propagating band as shown in Fig.~\ref{multibound}(a). For an arbitrary non-zero atom-waveguide $J$, {the lower energy level is always separated and gradually departs from the boundary of the propagating band. Meanwhile, the photonic distribution as shown in Fig.~\ref{multibound}(b) demonstrates that the photon is bounded
in the whole regime of the giant atom.} However, the property of the upper state depends on
the value of $J$. For small $J$, the upper energy level coincides with the edge of
the propagating band and {keeps flat, as shown in Fig.~\ref{multibound}(a). The corresponding photonic probability distribution} is shown in Fig.~\ref{multibound}(c). It shows that the {probability} for finding a photon in the resonator increases as it moves far away from the {giant atom}  for $J/\xi=1$ and $J/\xi=2$. It implies that the bound state with energy higher than the propagating band does not exist for small $J$.  Only for a larger $J$, this energy level will be separated from the {propagating band}, and the state becomes a bound one. However, the photonic distribution is still different from the lower one in that the photon mainly distributes at the two ends of the {giant atom}, and decays outside the {atom} regime as shown by the curve for $J/\xi=3$ in Fig.~\ref{multibound}(c). We note that the similar ``quantum phase transition" behavior associated with the upper bound state also exists in the waveguide setup, which couples to a small three-level atom~\cite{LQ2019}.

\begin{figure}
\begin{centering}
\includegraphics[width=\columnwidth]{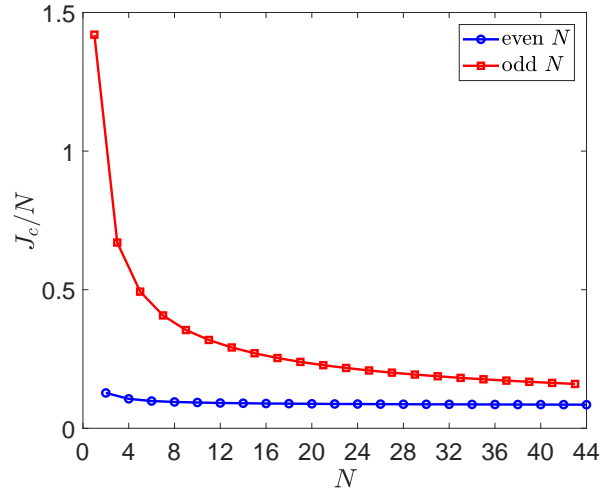}
\par\end{centering}
\caption{{The critical coupling strength $J_c$ as a function of the size of the giant atom for Case (II). The parameters are set as $\omega_c=\Omega=20\xi$.}}
\label{criticalstrength}
\end{figure}

{At last, we emphasize two points about the quantum phase transition. First, the phase transition only occurs for the upper state  but not the lower one in energy. To show this fact,  we plot the photonic probability distribution in Fig.~\ref{small}. It shows that, even for very small $J$ ($J/\xi=0.1$), the lower state is the bound state {for which the photon is localized around the
giant atom. For the upper state, we have shown the case for a waveguide with $101$ resonators (labelled by indexes $-50\rightarrow50$) under the periodical boundary condition and the giant atom couples to the resonators of $0-5$. It shows that the photon is repulsed away from the atom regime, and is a little localized at the $\pm50$ resonators, but the width is much larger than that of the lower bound state. To differ from the scattering states as discussed in Sec.~\ref{tscattering} and Sec.~\ref{mscattering}, we name the upper state below the critical point as extended state.} Second, in Fig.~\ref{multibound}(a), we have shown the quantum phase transition behavior when the giant atom couples to the resonators with index $j=0-N$ for $N=5$. In Fig.~\ref{criticalstrength}, we further exhibit the critical coupling strength $J_c$, at which the quantum phase transition occurs, as a function of the size of the giant atom $N$. We find that the behavior of {$J_c/N$} can be divided into two groups with odd and even $N$, respectively. In each group, {$J_c/N$ decreases with $N$ for small $N$ and tends to be constance. That is, the $J_c$ is approximately proportional to $N$ when $N$ is large enough.  Furthermore, the curve for odd $N$ is above that for even $N$, at least for $N\leq 45$ in our consideration.} We also note that, for $N=1$, it belongs to both of Case (I) and Case (II), we have discussed the single-photon scattering for this case in the Sec.~\ref{tscattering} and  it also shows a phase transition behavior as discussed in this section.}

\section{Conclusion}
\label{con}
In this paper, we have studied the single-photon scattering and bound states in a one-dimensional coupled-resonator waveguide with a dressed giant atom. The giant atom can couple to the waveguide via two or multiple resonators, and the {photon can propagate back and forth in the regime of giant atom. The interference effect will lead to a Breit-Wigner or Fano line shape for the reflection rate, depending on the size of the giant atom}.  {When the atom couples to two nonadjacent resonators, the energies for the bound state lie symmetrically outside the propagating band for arbitrary atom-waveguide coupling strength. When the atom couples to all of the resonators it covers,} we find the phase transition phenomenon based on the bound states. That is, the upper bound state in which the photon locates at the end of the giant atom only exists when the atom-waveguide coupling strength surpasses a critical value.

{We also note that, the phase transition is usually a genuine phenomenon in infinite-size system.
However, limited by our computing capacity, we only show the results in Fig.~\ref{criticalstrength} for the waveguide with $1201$ sites. Even for such finite-size system, we can also
observe a phase transition phenomenon, and we believe that the scaling behavior of $J_c\propto N$ ($N$ characterizes the size of the giant atom), which is implied by Fig.~\ref{criticalstrength}, will also work well for infinite-size system.}

We hope that our study will be applicable to quantum acoustics~\cite{YW2017,BA2018}, where the size of the emitter can be comparable to the wavelength of the phonons, and serves as a controller to the propagation of the phonons. Moreover, the bound state between an emitter and its non-Markovian environment is shown to be useful in for example preserving quantum entanglement~\cite{an2010}, suppressing dissipation~\cite{an2017}, enhancing quantum metrology~\cite{an2019}. Therefore, it is also interesting to discuss the application of the bound states in our discussions, here the waveguide actually serves as a structured non-Markovian environment for the giant atom. We also hope the interference effect for the propagation of the photon in the regime of the giant atom will be applicable in quantum control and quantum information processing.

\begin{acknowledgments}
 We thank J. L. Li for his help on figure polishing. This work is supported by  National Natural Science Foundation of China (Grant No. 11875011); Educational Commission of Jilin Province of China (Grant No. JJKH20190266KJ).
\end{acknowledgments}
\appendix%\appendixpage
\addcontentsline{toc}{section}{Appendices}\markboth{APPENDICES}{}
\begin{subappendices}
\section{Sing-photon bound states}
\label{A1}
In Eqs.~(\ref{eqtwobound}) and (\ref{eqmbound}), we have listed the transcendental equations for the energies of the bound states, when the {giant atom} couples to the waveguide via two and multiple resonators, respectively. Here, we will give the detailed derivation by transforming to the momentum space.

First, for the case in which the {giant atom}  couples to the waveguide via two resonators, the Hamiltonian in the momentum space is expressed as
\begin{equation}
H_{1}=\sum_{k}\omega_k a_{k}^{\dagger}a_{k}+\Omega|e\rangle\langle e|+\frac{J}{\sqrt{N_{0}}}\sum_{k}[a_{k}^{\dagger}(1+e^{ikN})\sigma^{-}+{\rm H.c.}],
\label{AH1}
\end{equation}
where $a_k=\sum_{n=-\infty}^{\infty} \exp{(-ikn)}a_n/\sqrt{N_0}$, with $N_0$ the length of the waveguide, and the dispersion relation is given by $\omega_k=\omega_c-2\xi\cos k$.
In the single-excitation subspace, the wave function has the form
\begin{equation}
|\psi\rangle=(b\sigma^{+}+\sum_{k}c_{k}a_{k}^{\dagger})|G\rangle,
\label{wavek}
\end{equation}
then the Sch\"{o}dinger equation $H_{1}|\psi\rangle=E|\psi\rangle$
{will give the coupled equations
\begin{eqnarray}
 b(E-\Omega)&=&\frac{J}{\sqrt{N_{0}}}
 \sum_k c_{k}(1+e^{-ikN}),\\
(E-\omega_c+2\xi\cos k)c_{k}&=&\frac{J}{\sqrt{N_{0}}}(1+e^{ikN})b.
\end{eqnarray}
 As a result, eliminating $c_k$, we will obtain
 \begin{eqnarray}
 E-\Omega_c&=&\frac{J^{2}}{N_{0}}\sum_{k}\frac{2[1+\cos(kN)]}{E-\omega_c+2\xi\cos(k)}\nonumber\\
&=&\frac{J^{2}}{\pi}\int_{-\pi}^{\pi}dk\frac{1+\cos(kN)}{E-\omega_c+2\xi\cos(k)},
\label{AEk}
\end{eqnarray}
which is Eq.~(\ref{eqtwobound}) in the main text.}

Second, we consider the case that the {giant atom}  couples to all of the resonators it covers in the waveguide, the Hamiltonian in the momentum space becomes
\begin{eqnarray}
H_{2}=&&\sum_{k}\omega_k a_{k}^{\dagger}a_{k}+\Omega|e\rangle\langle e|\nonumber \\&&+\frac{2J}{\sqrt{N_{0}}(N+1)}\sum_{n=0}^{N}\sum_{k}
[a_{k}^{\dagger}e^{ikn}\sigma^{-}+{\rm H.c.}].
\end{eqnarray}

{Reduplicating the process from Eq.~(\ref{wavek}) to (\ref{AEk}), we will end with
\begin{equation}
E-\Omega=\frac{2J^{2}}{(N+1)^{2}\pi}\int_{-\pi}^{\pi}dk
\frac{\sin^{2}[\frac{k(N+1)}{2}]}{\sin^{2}(\frac{k}{2})[E-\omega_c+2\xi\cos(k)]},
\end{equation}
which is Eq.~(\ref{eqmbound}) in the main text.}
\end{subappendices}

\end{document}